\documentclass[12pt]{article}
\usepackage{amsmath,amsfonts,amssymb,amsthm,textcomp}
\usepackage{amsthm}
\usepackage[font=footnotesize,labelsep=period]{caption}
\usepackage{pdfsync}
\usepackage{authblk}

\numberwithin{equation}{section}

\theoremstyle{plain}
\newtheorem{theorem}{Theorem}

\theoremstyle{definition}

\newcommand{\gen}[1]{\frac{\partial}{\partial{#1}}}

\newcommand{\pr}[1]{{\rm pr}^{(#1)}}
\newcommand{\curl}[1]{ \left\{#1\right\} }
\newcommand{\lie}{\mathfrak g}

\DeclareMathOperator{\Sl}{sl}

\DeclareMathOperator{\SL}{SL}

\DeclareMathOperator{\So}{so}

\DeclareMathOperator{\Vect}{Vect}

\begin{document}

\title{Analysis of the symmetry group and exact solutions of the  dispersionless KP
equation in $n+1$ dimensions}

\author[1]{J.~M. Conde\thanks{jconde@usfq.edu.ec}}
\author[2]{F.~G\"ung\"or\thanks{gungorf@itu.edu.tr}}
\affil[1]{Universidad San Francisco de Quito (USFQ), Quito, Ecuador
Departamento de Matem\'{a}ticas, Colegio de Ciencias e Ingenierias}
\affil[2]{Department of Mathematics, Faculty of Science and
Letters, Istanbul Technical University, 34469 Istanbul, Turkey}


\maketitle

\begin{abstract}
The Lie algebra of the symmetry group of the $(n+1)$-dimensional ge\-ne\-ra\-li\-zation of the dispersionless Kadomtsev--Petviashvili (dKP) equation is obtained and identified as a semi-direct sum of a finite dimensional simple Lie algebra and an infinite dimensional nilpotent subalgebra. Group transformation properties of solutions under the subalgebra $\Sl(2,\mathbb{R})$ are presented.  Known explicit analytic solutions in the literature are shown to be actually group-invariant solutions  corresponding to certain specific infinitesimal generators of the symmetry group.
\end{abstract}

\vspace{1,5mm}


\section{Introduction}
The $(n+1)$-dimensional generalization of the dispersionless Kadomtsev--Petviashvili equation (in short, it will be referred to as the dKP$_n$ equation)
\begin{equation}\label{dKPn}
  E(x,\vec y,t)=\left(u_t+u u_x \right)_x+\Delta_{\bot}u=0, \quad
\Delta_{\bot}=\sum\limits_{i=1}^{n-1}\partial^2_{y_i},\quad n\ge 2, \quad x\in \mathbb{R}, \quad \vec y\in \mathbb{R}^{n-1},
\end{equation}
where $u$ is a real-valued function depending on $n+1$ variables, $u=u(x,\vec y,t)=u(x,y_1,...,y_{n-1},t)$, was studied some time ago from several perspectives.
This equation describes the propagation of weakly nonlinear quasi one dimensional waves in
$n+1$ dimensions. Eq. \eqref{dKPn}, which is the $x$-dispersionless limit of the $n+1$ dimensional generalization of the Kadomtsev--Petviashvili (KP) equation is comprised of two competing terms; nonlinear term $uu_x$ and the $\Delta_{\bot}$ term (the Laplacian in the transverse variables), describing  diffraction in the transversal $(n-1)$-dimensional hyperplane   and includes as particular cases the integrable Riemann equation for $n=1$ (dKP$_1$), the integrable dispersionless KP (dKP), also called the Khokhlov--Zabolotskaya (KZ) equation, for $n=2$ (dKP$_2$) \cite{ZabolotskayaKhokhlov1969} and the nonintegrable KZ equation for $n=3$ (dKP$_3$). The potential Khokhlov--Zabolotskaya equation or Lin--Reissner--Tsien equation
$$2v_{xt}+v_xv_{xx}+v_{yy}=0$$
is related to the dKP$_2$ equation with the transformation $u=v_x$ and the scaling $t\to 2t$. The dKP$_3$ equation describes the propagation of a confined three-dimensional sound beam in a slightly non--linear medium without absorption or dispersion.
We refer to \cite{ManakovSantini2011} and \cite{SantucciSantini2016} for more details on physical motivation.

In a recent work  \cite{Sergyeyev2018}, a new systematic method  was introduced for construction of $3+1$-dimensional dispersionless systems using nonisospectral Lax pairs that involve contact vector fields. Within this approach,  a (3 + 1)-dimensional integrable generalization of the potential dKP equation (or the dKP$_3$ equation under the differential substitution $v=u_x$) was constructed
\begin{equation}\label{3+1-dKP}
  (4u_t+6u_x^2)_x-3(u_{yy}+u_{zz})=0.
\end{equation}

We organize this paper as follows.  In Section \ref{S2}, we obtain the Lie point symmetry algebra $\lie$ of dKP$_n$ equation \eqref{dKPn}  and identify its Lie algebraic structure. We write it  in a suitable basis with commutation relations. In Section \ref{S3}, we study  group transformations mapping solutions amongst themselves. This permits simple solutions like constants or more nontrivially $\vec{y}$-independent solutions, namely $(t,x)$-dependent solutions of dKP$_1$, to be mapped to new solutions in all independent variables $(t,x,\vec{y})$, depending on parameters and arbitrary functions of time.  We also derive solutions invariant under infinite dimensional subalgebras for appropriately chosen arbitrary functions figuring in the subalgebra, by reducing  the dKP$_n$ equation \eqref{dKPn} to the dKP$_1$ equation for which the Cauchy problem is implicitly solved for some given initial data. The obtained group invariant solutions  are then compared with the existing exact solutions, confirming that they all coincide.

\section{Lie point symmetries}\label{S2}

In this Section we apply the classical Lie algorithm  to find the Lie algebra $\lie$ of the  symmetry group or simply the symmetry algebra of Eq. \eqref{ZKSYM}.
We can write a general element of $\lie$ as a vector field
\begin{equation}\label{VF}
X = \tau \frac{\partial}{\partial t} +
\xi \frac{\partial}{\partial x} +
\sum_{i=1}^{n-1}\xi_i \frac{\partial}{\partial y_i} +
\eta \frac{\partial}{\partial u},
\end{equation}
where  $\tau$, $\xi$, $\xi_1$, $\xi_2$, \ldots, $\xi_{n-1}$ and $\eta$ are the coefficients of the vector field defined on the jet space $J(\mathbb{R}^{n+1},\mathbb{R})$ which has the local coordinates $(t,x,y_1,\ldots, y_{n-1},u)$. The algorithm consists of requiring that the second prolongation  $\pr{2}X$ of the vector field $X$ to the second order jet space $J^{2}(\mathbb{R}^{n+1},\mathbb{R})$ of the independent and dependent variables and all derivatives of the dependent variable up to order two   annihilates Eq. \eqref{ZKSYM} on its solution surface
\begin{equation}\label{sym-cond}
  \pr{2}X(E){\Big\vert}_{E=0}=0,
\end{equation}
where $E=0$ is the equation under study.

Applying the condition \eqref{sym-cond} and splitting the resulting expression with respect to the linearly independent derivative terms we obtain an overdetermined system of  determining equations (first order linear PDEs). We solve this determining  system and express the general element of the symmetry algebra as a theorem:
\begin{theorem}
The Lie symmetry algebra $\lie$ of \eqref{ZKSYM} for $n>2$ is realized by vector fields \eqref{VF} with coefficients given by
\begin{eqnarray}
&&
\xi=\frac{\tau_2}{n+7}\left[2(5-n)xt-3\sum_{i=1}^{n-1} y_i^2\right]+(2\sigma-\tau_1)x-\frac{1}{2}\sum_{i=1}^{n-1} F_i'(t) y_i+F_n(t),
\nonumber\\
&&\tau=\tau_2t^2+\tau_1 t+\tau_0,
\nonumber\\
\end{eqnarray}
\[
\vec{\xi}=\left( \begin{array}{cccc}
 \xi_1\\
 \xi_2 \\
 \vdots  \\
 \xi_{n-1}
\end{array} \right)
=
\left( \begin{array}{cccc}
 \delta & c_{12} & c_{13} & c_{1,n-1}  \\
 -c_{12} & \delta & c_{23} & c_{2,n-1} \\
 -c_{13} & -c_{23} & \delta & c_{3,n-1} \\
 \vdots & \vdots & \ddots & \vdots \\
 -c_{1,n-1} &  -c_{2,n-1} & \cdots & \delta
\end{array} \right)
\left( \begin{array}{cccc}
 y_1\\
 y_2 \\
 \vdots  \\
 y_{n-1}
\end{array} \right)
+
\left( \begin{array}{cccc}
 F_1(t)\\
 F_2(t)\\
 \vdots  \\
 F_{n-1}(t),
\end{array} \right),
\]
with
\begin{equation*}
\delta=\frac{12\tau_2 t}{n+7}+\sigma,
\end{equation*}

\begin{equation*}
\eta=\frac{1}{n+7}
\left[-4(n+1)\tau_2 t +2(n+7)(\sigma - \tau_1 )\right]u +\frac{2(5-n)}{n+7}\tau_2 x-\frac{1}{2}\sum_{i=1}^{n-1} F_i''(t) y_i+F_n'(t), \end{equation*}
where $\tau_2$, $\tau_1$, $\tau_0$, $\sigma$, $c_{ij}=-c_{ji}$ for $i,j=1, ... , n-1$  are arbitrary constants, and $F_i(t)$, for $i=1, \ldots , n-1$ are arbitrary functions of time. $\vec{\xi}$ can be written in the form $\vec{\xi}=\mathbf{C}\vec{y}+\delta \mathbf{I}+\vec{F}(t)$, $\vec{y}=(y_1,y_2,\ldots,y_{n-1})^{T}$, $\vec{F}(t)=(F_1(t),\ldots, F_{n-1}(t))^{T}$, $\mathbf{I}$ is the identity matrix of size $(n-1)\times(n-1)$ and $\mathbf{C}$ is a skew-symmetric matrix of the same size.

\end{theorem}

A basis for the symmetry algebra $\lie$ is given by the following vector fields
\begin{equation}\label{basis}
\begin{gathered}
  T=\gen t,  \quad \tilde{D}=t\gen t-x\gen x-2u \gen u, \\
  C=t^2\gen t+\frac{1}{n+7}[2(5-n)xt-3|\vec{y}|^2]\gen x+\frac{12t}{n+7}\vec{y}\cdot\gen {\vec{y}}+\\
  +\frac{2}{n+7}[(5-n)x-2(n+1)tu]\gen u,
\end{gathered}
\end{equation}
\begin{equation}\label{scaling}
  D_0=2x\gen x+\vec{y}\cdot\gen {\vec{y}}+2u\gen u,
\end{equation}
\begin{equation}\label{rot}
  R_{ij}=y_i\gen {y_j}-y_j\gen {y_i}, \quad i<j, \quad i,j=1,2,\ldots, n-1,
\end{equation}
\begin{equation}\label{radical}
\begin{split}
    & X(F_n)=F_n(t)\gen x+F_n'(t)\gen u, \\
     & Y(F_i)=F_i(t)\gen {y_i}-\frac{1}{2}F_i'(t)y_i\gen x-\frac{1}{2}F_i''(t)y_i\gen u, \quad i=1,\ldots, n-1,
\end{split}
\end{equation}
where
$$\vec{y}=(y_1,y_2,\ldots,y_{n-1}), \quad |\vec{y}|^2=\sum_{j=1}^{n-1}y_j^2,  \quad \vec{y}\cdot\gen {\vec{y}}=\sum_{j=1}^{n-1}y_j\gen {y_j}.$$ Here $T$, $\tilde{D}$, $D_0$, $C$ and $R_{ij}$ correspond to time translations, dilations,  projective transformations, and rotations on the $y_iy_j$-hyperplane, respectively. The commutation relations among $T,\tilde{D},C$ are
\begin{equation}\label{sl2-com}
  [T,\tilde{D}]=T,  \quad [T,C]=2\tilde{D}+D_0,  \quad [\tilde{D},C]=C.
\end{equation}
Taking into account that $D_0$ commutes with $T,\tilde{D},C$, together with  change of basis \begin{equation}\label{D}
  D=\tilde{D}+\frac{D_0}{2}=t\gen t+\frac{1}{2}\vec{y}\cdot\gen {\vec{y}}-u\gen u.
\end{equation}
we see that $\curl{T,D,C}$ realizes the $\Sl(2,\mathbb{R})$ algebra  with the commutation relations
\begin{equation}\label{sl2-com-2}
  [T,D]=T,  \quad [T,C]=2D,  \quad [D,C]=C.
\end{equation}
The  commutation relations among other elements are
\begin{subequations}\label{com}
\begin{gather}
[D_0, X(G)]=-2X(G),  \quad [D_0, Y(F_i)]=-Y(F_i), \quad [X(G),Y(F_i)]=0,\\
[X(G),T]=-X(G'),  \quad [X(G),D]=-X(tG'),  \quad [X(G),C]=-X(t^2G'),\\
[Y(F_i),T]=-Y(F'_i),  \quad [Y(F_i),D]=Y(\frac{F_i}{2}-tF_i'),  \quad [Y(F_i),C]=Y(tF_i-t^2F_i'),\\
[Y(F_i),Y(G_j)]=-\frac{1}{2}\delta_{ij}X(F_iG_j'-F_i'G_j),\\
[R_{ij},X(G)]=0,  \quad [R_{ij},Y(F_k)]=-\delta_{kj}Y_j(F_i),
\end{gather}
where we have defined $$Y_j(F_i)\equiv F_i(t) \gen {y_j}-\frac{1}{2}F_i'(t)y_j\gen x-\frac{1}{2}F_i''(t)y_j\gen u,  \quad Y_j(F_j)\equiv Y(F_j).$$
\end{subequations}
Rotations $R_{ij}$ commute with $T,D,C$.

The point Lie symmetry algebra $\lie$ of Eq. \eqref{dKPn} is infinite-dimensional and
can be written as  a semi-direct sum Lie algebra (Levi decomposition), $\lie=S \uplus R,$ where $S=\Sl(2,\mathbb{R})\oplus \So(n-1)$ is the finite-dimensional semisimple part, namely Levi factor
of $\lie$ and $$R = \curl{X(F_n), Y(F_1), \ldots, Y(F_{n-1}),  D_0}$$ is its infinite-dimensional radical (nonnilpotent ideal). Here $\So(n-1)$ is the $(n-1)(n-2)/2$-dimensional algebra of rotations in $\mathbb{R}^{n-1}$. The subalgebra $R$ has the  structure of a centerless Kac-Moody algebra.

In the special case when $n=2$, the equation is completely integrable and the structure of the
symmetry algebra $\lie$ is exceptionally different than $n>2$. A  basis for $\lie$ ~\cite{Schwarz1987} is given by
\begin{equation}\label{ZKSYM}
\begin{split}
& {T} (f) = f \gen t + \frac{1}{6} \left(2 x f' - y^2
f''\right) \gen x + \frac{2 y}{3}  f' \gen y + \frac{1}{6}
\left(- 4 u f' + 2 x f'' - y^2 f'''\right) \gen u,\\
& {X}(g) = g \gen x + g' \gen u, \\
& {Y}(h) =  h\gen y-\frac{1}{2} y h'\gen x -\frac{1}{2} y
h'' \gen u, \\
& D = 2 x \gen x + y \gen y + 2 u \gen u ,
\end{split}
\end{equation}
where $f, g, h$ are arbitrary smooth functions of the time variable $t,$
and the prime denotes  derivative with respect to $t$. Note that the generators obtained by restricting $f$ to $1,t,t^2$
$$T(1)=T,  \quad T(t)=\tilde{D}+\frac{2}{3}D_0,  \quad T(t^2)=C,$$  realize  the $\Sl(2,\mathbb{R})$ algebra as a finite-dimensional subalgebra, where $T, \tilde{D}, D_0, C$ are as defined in \eqref{basis} and \eqref{scaling} with $n=2$.
The commutation relations are
\begin{alignat*}{2}
[{T}(f_1), {T}(f_2)]\; &= \; {T}{ (f_1 f_2'-f_1' f_2 )},\\
[D, {X}(g)]\;  &= \; -2 {X}(g),\medspace &
[{X}(g), {Y}(h)]\; &= \;
0  \\
[D, {Y}(h)]\;  &= \; - {Y}(h),\medspace  &
[{X}(g),{T}(f)]\;
&= \; {X}{(\frac{1}{3} f' g - f g')},   \\
[D, {T}(f)]\; &= \;0, \medspace & [{Y}(h), {T}(f)]\; &= \; {Y}{ (\frac{2}{3} f'h - f h')}, \\
[{X}(g_1), {X}(g_2)] \; &= \; 0,  \medspace &  [{Y}(h_1), {Y}(h_2)] \; &= \; \frac{1}{2}{X}{ (h_1 h_2'- h_1' h_2 )}.
\end{alignat*}

The  symmetry algebra $\lie$
can be written as  a semi-direct sum Lie algebra (Levi decomposition), $\lie=S \uplus R,$ where $S=
\{T(f)\}$ is the (infinite-dimensional) semisimple part, also called Levi factor
of $\lie$ and $R = \{X(g), Y(h), D\}$ is its radical.    $S$  is a
simple Lie algebra, i.e. it has no nontrivial ideal.   The radical
$R$ (maximal solvable ideal) is actually nonnilpotent.  The algebra $S$ can be identified as a centerless Virasoro algebra (the Witt algebra).  The algebra $R$ is a subalgebra of a centerless Kac-Moody algebra \cite{Gungor2010a}.
The vector fields $T(f)$ generate a Lie algebra isomorphic to the algebra $\Vect(\mathbb{R})$ of real vector fields on $\mathbb{R}$ . If the parameter $t$ is compactified, we can replace $\Vect(\mathbb{R})$ by the Witt algebra $\Vect(\mathbb{S}^1)$. The presence of $S$ implies that the equation is invariant under arbitrary reparametrizations of time.

This typical symmetry structure arises  in completely integrable evolutionary equations in $2+1$-dimensions like KP equation \cite{DavidKamranLeviWinternitz1985, DavidKamranLeviWinternitz1986}, three-wave resonance equations \cite{MartinaWinternitz1989}, Davey-Stewertson system \cite{ChampagneWinternitz1988} and a few others (see \cite{Gungor2006, Basarab-HorwathGuengoerOezemir2013} for a further discussion). Indeed, this is the case here. Eq. \eqref{ZKSYM} is known to enjoy all the features of integrability. When $n>2$ its integrability is destroyed and the Virasoro algebra reduces to its finite dimensional subalgebra, which is $\Sl(2,\mathbb{R})$. The infinite-dimensional subalgebra $R$ remains intact.

We comment that in this case $\Sl(2,\mathbb{R})$ subalgebra can be embedded into the centerless Virasoro algebra (the Witt algebra)
generated by the algebra of vector fields $\curl{L_n=T(t^{n+1}): n\in \mathbb{Z}}$, whose elements satisfy the commutation relations
$$[L_m,L_n]=(n-m)L_{m+n}.$$

Group-invariant solutions of \eqref{ZKSYM} can be found in Ref. \cite{Ndogmo2008, Ndogmo2009}.

We remark that in \cite{GungorOzemir2015} the following variable coefficient version was studied
\begin{equation}\label{gdKP}
  (u_t+p(t)uu_x)_x+\sigma(t)u_{yy}=0.
\end{equation}
The conditions on the coefficients were established to ensure the transformability to its constant coefficient case. The Virasoro structure only survives under this equivalence.

\section{Group Transformations and exact solutions}\label{S3}
The subgroups corresponding to $T$ and $D$ are easy to obtain. They are simply translations in $t$: $t\to t+t_0$ and dilations: $(t,x,y_i,u)\to (\lambda t, x, \lambda^{1/2}y_i,\lambda^{-1}u)$, $\lambda>0$, where $t_0$, $\lambda$ are the group parameters. The one-parameter  projective symmetry group generated by  the subalgebra $\curl{C}$ (the flow of the vector field $C$),  $G_{\varepsilon}:(\tilde{t},\tilde{x},\tilde{y}_i,\tilde{u})=\exp(\varepsilon C).(t,x,y_i,u)$    is  obtained by integrating the vector field $C$ as follows:
\begin{equation}\label{expC}
  \begin{split}
& \frac{d\tilde{t}}{d\varepsilon}=\tilde{t}^{2},\\
& \frac{d\tilde{x}}{d\varepsilon}=(n+7)^{-1}[2(5-n)\tilde{t}\tilde{x}-3|\tilde{\vec{y}}|^2],\\
&\frac{d\tilde{\vec{y}}}{d\varepsilon}=12(n+7)^{-1}\tilde{t}\tilde{\vec{y}},\\
&\frac{d\tilde{u}}{d\varepsilon}=2(n+7)^{-1}[(5-n)\tilde{x}-2(n+1)\tilde{t}\tilde{u}],
  \end{split}
\end{equation}
subject to the initial conditions
$$\tilde{t}(t,x,\vec{y};\varepsilon)\big\vert_{\varepsilon=0}=t,  \quad \tilde{x}(t,x,\vec{y};\varepsilon)\big\vert_{\varepsilon=0}=x,  \quad \tilde{\vec{y}}(t,x,\vec{y};\varepsilon)\big\vert_{\varepsilon=0}=\vec{y}, \quad \tilde{u}(t,x,\vec{y};\varepsilon)\big\vert_{\varepsilon=0}=u.$$
We find
\begin{equation}\label{group-proj}
  \begin{split}
    &\tilde{t}=\frac{t}{1-\varepsilon t},  \quad \tilde{x}=(1-\varepsilon t)^{-\alpha}\left(x-\frac{3\varepsilon |\vec{y}|^2}{(n+7)(1-\varepsilon t)}\right),  \quad \tilde{y}_i=(1-\varepsilon t)^{-12/(n+7)}y_i, \\
  &\tilde{u}(t,x,\vec{y};\varepsilon)=(1-\varepsilon t)^{\beta}\left\{u+\alpha\left(\frac{\varepsilon x}{1-\varepsilon t}-\frac{3\varepsilon^2|y|^2}{2(n+7)(1-\varepsilon t)^2}\right)\right\},  \end{split}
\end{equation}
where $\varepsilon$ is the group parameter and $\alpha$, $\beta$ are defined by
$$\alpha=\frac{2(5-n)}{n+7},  \quad \beta=\frac{4(n+1)}{n+7}.$$
The  Lie group of local point transformations \eqref{group-proj} transforms a solution $u(t,x,\vec{y})$  into a new solution $\tilde{u}(\tilde{t},\tilde{x},\vec{\tilde{y}})=G_{\varepsilon}\circ u(t,x,\vec{y})$ (see formula \eqref{new-sol}).

In the special case of $n=2$ the symmetry transformations can be expressed as
\begin{equation}\label{sym-group-Vir}
  \begin{split}
    &\tilde{t}=T(t), \quad \tilde{x}=T'(t)^{1/3}x-\frac{1}{6}T''(T')^{-2/3}y^2, \quad \tilde{y}=T'(t)^{2/3}y, \\
      & \tilde{u}=T'(t)^{-2/3} \Bigl\{ u+\frac{T''(t)}{3T'(t)}x-\frac{1}{6}\Bigl[\curl{T;t}+\frac{1}{6}
      \Bigl(\frac{T''}{T'}\Bigr)^2\Bigr]y^2\Bigr\},
  \end{split}
\end{equation}
where $T(t)$ is a arbitrary smooth function of $t$ and $\curl{T;t}$ is the Schwarzian derivative of $T$
$$\curl{T;t}=\frac{T'''}{T'}-\frac{3}{2}\left(\frac{T''}{T'}\right)^2.$$
A group parameter $\lambda$ can be introduced into \eqref{sym-group-Vir} through the transformation $$\tilde{t}=T(t;\lambda)=F^{-1}(F+\lambda)(t), \quad F(t)=\int^{t}\frac{ds}{f(s)},$$
with the property $T(t;0)=t$ and $T_{\lambda}(t;0)=f(t)$. Here
$f(t)$ is the $t$-coefficient of $T(f)$ in \eqref{ZKSYM} and  $F^{-1}(t)$ is the inverse to $F(t)$. The corresponding infinite-dimensional symmetry group (Virasoro group)  is the exponentiation $\exp(\lambda T(f))$ of the vector filed $T(f)$. For the particular choice  $f(t)=t^2$, we have $T(t;\lambda)=t/(1-\lambda t)$ and the formula \eqref{sym-group-Vir} coincides with the projective transformation \eqref{group-proj} for $n=2$, $(\alpha,\beta)=(2/3,4/3)$. Taking $f(t)=t,1$ we have  $T(t;\lambda)=e^{{\lambda}}t$ and $T(t;\lambda)=t+\lambda$, generating dilation and translation symmetries, respectively.

If $T(t)$ in \eqref{sym-group-Vir} is restricted to the Moebius transformations $T(t)=(at+b)/(ct+d)$, $ad-bc=1$ for which $\curl{T;t}=0$, and $T''/T'=-2c/(ct+d)$, then the transformation formula \eqref{sym-group-Vir} is reduced to the $\SL(2,\mathbb{R})$ invariance group  of the equation depending on three group parameters $a,b,c$:
\begin{equation}\label{SL2}
 \begin{split}
    &\tilde{t}=\frac{at+b}{ct+d}, \quad \tilde{x}=(ct+d)^{-2/3}x+\frac{c}{3}(ct+d)^{-5/3}y^2, \quad \tilde{y}=(ct+d)^{-4/3}y, \\
      & \tilde{u}=(ct+d)^{4/3} \Bigl[ u-\frac{2cx}{3(ct+d)}-\frac{c^2y^2}{9(ct+d)^2}\Bigr].
  \end{split}
\end{equation}

The vector field $D_0$ generates the dilation group $(t,x,y_i,u)\to (t, \lambda^2 x, \lambda y_i,\lambda^{2}u)$.

The vector field $X(F_n)$  generates the group transformations
\begin{equation}\label{gen-Galilei}
  \tilde{t}=t,  \quad \tilde{x}=x+\lambda F_n(t), \quad \tilde{y}_i=y_i, \quad \tilde{u}=u+\lambda F'_n(t).
\end{equation}
Physically it is a transformation to a frame moving with an arbitrary acceleration in the $x$ direction.
For $F_n=\mbox{const}$ we have a translation and for $F_n=t$  a Galilei transformation. The solutions depending on the arbitrary function $F_n(t)$ are obtained from the reduced equation
$$\Delta_\bot F(t,\vec{y})=-\frac{F''_n(t)}{F_n(t)}$$ by the transformation formula
$$u=\frac{F'_n(t)}{F_n(t)}x+F(t,\vec{y}).$$

The  vector fields $Y(F_1),\ldots, Y(F_{n-1})$ generate the symmetry transformations
\begin{equation}\label{expY}
  \tilde{t}=t,  \quad \tilde{x}=x+\sum_{i=1}^{n-1}(\lambda_i y_i F'_i-\lambda_i^2F_iF_i'), \quad \tilde{y}_j=y_j-2\lambda_j F_j,  \quad \tilde{u}=u+\sum_{i=1}^{n-1}(\lambda_iF_i''y_i-\lambda_i^2F_iF_i''),
\end{equation}
where $j=1,\ldots, n-1$ and  $\lambda$'s are the group parameters.

In particular, the choice $F_i(t)=t$ leads to the transformation
\begin{equation}\label{par-Y}
  \tilde{t}=t,  \quad \tilde{x}=x+\sum_{i=1}^{n-1}(\lambda_i y_i -\lambda_i^2t), \quad \tilde{y}_j=y_j-2\lambda_j t,  \quad \tilde{u}=u.
\end{equation}
This is the transformation used in \cite{ManakovSantini2011} to construct group invariant solutions, namely those mapped into themselves by the subgroup \eqref{par-Y}. The corresponding infinitesimal generators have the form of quasi-rotation
\begin{equation}\label{Y-t}
 Y_i(t)=2t\gen{y_i} - y_i\gen x, \quad i=1,2,\ldots, n-1.
\end{equation}
In order to perform symmetry reduction, we need the invariants of generators \eqref{Y-t}. They are  $t$, $u$ and  $\xi=x+(4t)^{-1}|\vec{y}|^2$ (a paraboloid). Solutions invariant under $Y_j(t)$ are obtained by  the reduction formula $u=v(t,\xi)$, where $v$ satisfies
\begin{equation}\label{hopf-0}
  v_t+\frac{n-1}{2t}v+vv_{\xi}=0.
\end{equation}
This equation is equivalent to the Riemann-Hopf equation $q_{\tau}+qq_{\xi}=0$ by the change of variable $v=a(t)q(\tau,\xi)$ with $a(t)$ and $\tau(t)$ appropriately chosen (see \cite{ManakovSantini2011}). This means that the authors of \cite{ManakovSantini2011} rotated the general solution of $u_t+uu_x=0$ into solution of \eqref{dKPn}. For the sake of completeness, we reproduce the invariant (implicit)  solution found in \cite{ManakovSantini2011}
\begin{equation}\label{exact-dKPn}
u=\left\{
\begin{array}{ll}
\displaystyle
t^{-\frac{n-1}{2}}F\left(x+\frac{1}{4t}\sum\limits_{i=1}^{n-1}y^2_i-\frac{2ut}{3-n}\right), &  n\ne 3, \\[.3cm]
\displaystyle t^{-1}F\left(x+\frac{1}{4t}\sum\limits_{i=1}^{n-1}y^2_i-u~t\ln t \right), & n=3 ,
\end{array}
\right.
\end{equation}
where $F$ is an arbitrary function of one argument.

One can of course use  general invariants of the transformations \eqref{expY} for arbitrary $F_i$ to perform a similar reduction. They are given by $t$ and the following two
\begin{equation}\label{inv}
  \xi=x+\frac{1}{4}\sum_{i=1}^{n-1}\frac{F'_i}{F_i}y_i^2,  \quad v=u+\frac{1}{4}\sum_{i=1}^{n-1}\frac{F''_i}{F_i}y_i^2\equiv u-R(t,\vec{y})
\end{equation}
Substitution of the symmetry ansatz
\begin{equation}\label{sym-ansatz}
  u=v(t,\xi)+R(t,\vec{y})
\end{equation}
into \eqref{dKPn} gives the invariant equation
\begin{equation}\label{red-1}
  (v_t+vv_{\xi})_{\xi}+(\xi_t+(\nabla \xi)^2+R)v_{\xi\xi}+(\Delta_\bot \xi)v_{\xi}+\Delta_\bot R=0.
\end{equation}
The coefficient of $v_{\xi\xi}$ vanishes and the reduced equation simplifies to
\begin{equation}\label{red-2}
  (v_t+vv_{\xi}+A(t)v)_{\xi}+B(t)=0,
\end{equation}
where we have defined
$$A(t)\equiv\Delta_{\bot} \xi=\frac{1}{2}\sum_{i=1}^{n-1}\frac{F'_i}{F_i}, \quad  B(t)\equiv\Delta_{\bot} R=-\frac{1}{2}\sum_{i=1}^{n-1}\frac{F''_i}{F_i}.$$
The arbitrary functions $F_i(t)$ can always be arranged such that $B(t)=0$. We assume that this is the case here.
Eq. \eqref{red-2} can then be integrated once to give
\begin{equation}\label{red-3}
 v_t+vv_{\xi}+A(t)v+g(t)=0,
\end{equation}
where $g(t)$ is an arbitrary function. The transformation
\begin{equation}\label{a-tau}
  v=a(t)q(\tau,\xi),  \quad a(t)=\exp{\left[-\int A(t)dt\right]},  \quad \tau=\int a(t)dt
\end{equation}
can be used to set $A(t)$ equal to zero in \eqref{red-3}. Furthermore, $g(t)$ can be removed by a time dependent translation of $\xi$ and $v$ (see Eq. \eqref{tr-f20}). The reduced equation is again $q_\tau+qq_{\xi}=0$.

The special choice $F_i(t)=t-t_i$, with $t_i$ being arbitrary constants, leads to
\begin{equation}\label{gen-sim}
  \xi=x+\frac{1}{4}\sum_{i=1}^{n-1}\frac{y_i^2}{t-t_i},  \quad A(t)=\frac{1}{2}\sum_{i=1}^{n-1}\frac{1}{t-t_i},  \quad B(t)=0
\end{equation}
and the invariant equation
\begin{equation}\label{red-4}
  v_t+vv_{\xi}+A(t)v=0.
\end{equation}
The functions $a(t)$ and $\tau(t)$ in the transformation \eqref{a-tau} are found from the relations
\begin{equation}\label{a-tau-special}
  a(t)=\prod_{i=1}^{n-1}(t-t_i)^{-1/2},  \quad \dot{\tau}(t)=a(t)
\end{equation}
so that $v=a(t)q(\tau,\xi)$ takes Eq. \eqref{red-4} to $q_{\tau}+qq_{\xi}=0$.

In particular, for $t_i=0$, $i=1,2,\ldots, n-1$, we have $A(t)=(n-1)/(2t)$, $B(t)=0$, $a(t)=t^{-(n-1)/2}$ and Eq. \eqref{hopf-0} is recovered. Different special choices of $F_i$ lead to different reductions and invariant solutions.

Motivated by the form of the group invariant solutions of \cite{ManakovSantini2011},   the authors of \cite{KamchatnovPavlov2016} used an ansatz to study reductions of the generalized DKPn equation
\begin{equation}\label{gen-dKPn}
  (u_t+u^Nu_x)_x+\frac{1}{2}\Delta_{\bot}u=0,  \quad N\in\mathbb{Z^{+}}
\end{equation}
to the generalized Riemann--Hopf equation
\begin{equation}\label{reduced-eq}
  U_T+U^NU_X+mU=g(t),
\end{equation}
where $m$ is some constant and $g(t)$ is an arbitrary function. Up to the change of variables $(t,x,u)\to (\tau(t),x,a(\tau)u(\tau,x))$,  this reduction coincides with the symmetry reduction  corresponding to the subalgebras $Y_i(t-t_i)$.  Indeed, contrary to the comment in \cite{KamchatnovPavlov2016}, the exact solutions found there are due to the invariance with the arbitrary functions $F_i(t)$ figuring in the subalgebras of the symmetry algebra of the equation specialized. In this setting, in the particular case $N=1$ and $n=2,3$, $t_i=0$ ($i\in\curl{1,2}$),  the solutions invariant under \eqref{Y-t} were recovered as already discussed before. Also, the solution constructed in \cite{KamchatnovPavlov2016} with the choice $N=1$, $n=3$ (with the notation of the present paper) $t_1<0$, $t_2>0$, $g(t)=0$ is a group invariant (similarity or self-similar) solution.

We note that Eq. \eqref{gen-dKPn} still remains invariant under the symmetry group generated by the vector field \eqref{Y-t}, but not under $Y(F_i)$ unless $F''_i=0$, and this invariance has been used in Ref. \cite{SantucciSantini2016} to construct group invariant solutions of \eqref{gen-dKPn} by reducing to the generalized Riemann--Hopf equation $u_t+u^Nu_x=0$ having an  implicit general solution. The fact that the group invariant solutions can be expressed implicitly depending on  an arbitrary function allows to solve a Cauchy problem  for a small initial condition.

Finally, the generators $R_{ij}$ correspond to the invariance of the transversal Laplace operator $\Delta_\bot$ under rotations. The invariant solutions will be of the form $u=v(t,x,|\vec{y}|)$.

The symmetry group \eqref{group-proj} can be used to transform trivial solutions like a constant solution or $\vec{y}$ independent solutions to other solutions depending on all variables. $\vec{y}$ independent solutions satisfy the quasilinear Riemann--Hopf  equation
\begin{equation}\label{Riemann}
  u_t+uu_x=f(t),
\end{equation}
where $f$ is an arbitrary integration function to be determined from the initial condition. The following point transformation transforms away $f(t)$
\begin{equation}\label{tr-f20}
  \tilde{t}=t,  \quad \tilde{x}=x+\lambda(t), \quad \tilde{u}=u+g(t),
\end{equation}
with $\lambda$, $g$ satisfying $\ddot{\lambda}-f=0$ and $\dot{\lambda}-g=0$. So the general solution $u(t,x)$ of the first order nonlinear PDE $u_t+uu_x=0$ (also known as inviscid Burgers' equation) determined implicitly by
\begin{equation}\label{gen-sol}
  u=F(\xi)=F(x-tu),
\end{equation}
with $F$ being arbitrary differentiable function of the characteristic coordinate $\xi$,
will generate new solutions of the form $u(t,x,y_i)$ in implicit form depending on the parameter $\varepsilon$ and the arbitrary function $F$. We recall that  the solution $u$ breaks at $t=t_b$ on the characteristic $\xi_b$, where $t_b=-1/F'(\xi_b)>0$. In other words, if $F'(\xi_b)<0$ the solution suffers a gradient catastrophe type of singularity.

From \eqref{group-proj} we can state that if $U(T,X,\vec{Y})$ is a solution of \eqref{dKPn}, then so is
\begin{equation}\label{new-sol}
  u(t,x,\vec{y})=(1+\varepsilon t)^{-\beta}U(T,X,\vec{Y})-\frac{(n-5)}{(n+7)^2}(1+\varepsilon t)^{-2}\left[2(n+7)(1+\varepsilon t)x+3\varepsilon^2 |\vec{y}|^2\right],
\end{equation}
\begin{equation}\label{TXY}
  T=\frac{t}{1+\varepsilon t},  \quad X=(1+\varepsilon t)^{-\alpha}\left(x+\frac{3\varepsilon |\vec{y}|^2}{(n+7)(1+\varepsilon t)}\right),  \quad \vec{Y}=(1+\varepsilon t)^{-12/(n+7)}\vec{y}.
\end{equation}
In the special case $n=5$ ($\alpha=0,\beta=2$), the above solution formula substantially simplifies to the form
$$u=(1+\varepsilon t)^{-2}U(T,X,\vec{Y}),  \quad T=\frac{t}{1+\varepsilon t},  \quad X=x+\frac{\varepsilon}{4(1+\varepsilon t)}|\vec{y}|^2, \quad \vec{Y}=\frac{\vec{y}}{1+\varepsilon t}.$$

Action of the $SL(2,\mathbb{R})$ symmetry group on solutions can be formulated as follows:
\begin{equation}\label{new-sol-SL2}
\begin{split}
  u(t,x,\vec{y})=(ct+d)^{-\beta}U(T,X,\vec{Y})-\frac{(n-5)}{(n+7)^2}(ct+d)^{-2}\left[2(n+7)(ct+d)x+3c^2 |\vec{y}|^2\right], \\
  T=\frac{at+b}{ct+d},  \quad X=(ct+d)^{-\alpha}\left(x+\frac{3c |\vec{y}|^2}{(n+7)(ct+d)}\right),  \quad \vec{Y}=(ct+d)^{-12/(n+7)}\vec{y},
\end{split}
\end{equation}
where $ad-bc=1$, $a,b,c$ are the group parameters and $\alpha$, $\beta$ ($\beta>0$) were defined in \eqref{group-proj}.  We comment that this holds for any value of $n$. For $n=2$ there is a more general formula involving an arbitrary function rather than parameters (see the symmetry group \eqref{sym-group-Vir}).

Restriction of the arbitrary function $f(t)$ of \eqref{ZKSYM} to a quadratic function results in this formula.   This type of formulas \eqref{new-sol-SL2}  play an important  role in demonstrating existence of the blow-up profiles for some appropriate Cauchy problem using  some simple solutions $U$ (like $\vec{y}$ independent solutions among many others).

On the other hand the scaling symmetry $D_0$ implies that if $u(t,x,\vec{y})$ solves \eqref{dKPn}, then so does $\tilde{u}(t,x,\vec{y})=\lambda^2 u(t,\lambda^{-2}x,\lambda^{-1}\vec{y})$, $\lambda>0$. With further application of  this invariance we can obtain new solutions depending on four parameters ($a,b,c,\lambda$) from known solutions.

\bibliographystyle{unsrt}

\end{document}